\documentclass[aps,prb,twocolumn,superscriptaddress]{revtex4-2}

\usepackage{amsmath, amssymb}
\usepackage{xcolor} 
\usepackage{soul}
\usepackage{float}
\usepackage[colorlinks = true, urlcolor = cyan, linkcolor=blue, citecolor=blue, filecolor=magenta]{hyperref}
\usepackage{adjustbox}
\usepackage{booktabs}

\newcommand{\bra}[1]{\langle #1|}
\newcommand{\ket}[1]{|#1\rangle}
\newcommand{\braket}[3]{\left\langle#1\left|#2\right|#3\right\rangle}
\newcommand{\icol}[1]{\left(\begin{smallmatrix}#1\end{smallmatrix}\right)} 

\begin{document}

\title{Exact and Fixed-Point Grover Search with Qudits}

\author{Tanay Roy}
\email{roytanay@fnal.gov}
\affiliation{Superconducting Quantum Materials and Systems (SQMS) Center, Fermi National Accelerator Laboratory, Batavia, IL 60510, USA}


\date{\today}

\begin{abstract}
Grover's algorithm provides a quadratic speedup for searching unstructured databases and is traditionally implemented with qubits in Hilbert spaces whose dimensions are powers of two. With the advent of quantum platforms utilizing qudits---quantum systems with more than two levels---there is a need to generalize Grover search to these architectures, including heterogeneous systems with qudits of varying dimensions. Here, we present a unified framework for qudit-based Grover search, detailing the construction of oracles and diffusion operators with and without ancilla qubits and generalizing deterministic and fixed-point search variants that ensure exact or bounded success probabilities. We analyze phase-matching techniques and provide explicit circuit decompositions suitable for diverse hardware platforms. We also compare the corresponding trajectories on the Bloch sphere to provide an intuitive visualization of how the different phase choices amplify the target state. These results facilitate flexible, hardware-oriented protocols for implementing Grover search on qudit processors, potentially reducing circuit depth and enhancing success probabilities, thereby offering a practical toolkit for quantum computation and sensing applications leveraging multilevel quantum systems.

\end{abstract}

\maketitle

\section{Introduction}
In the landscape of quantum algorithms, Grover's algorithm~\cite{Grover_algo} stands as a seminal milestone, proving a clear quantum advantage: a marked item in an unsorted database of size $N$ can be found using only $\mathcal{O}(\sqrt{N})$ oracle queries, compared with the $\mathcal{O}(N)$ queries required classically. Because of this quadratic speedup and the intuitive geometric interpretation of its dynamics on the Bloch sphere, Grover search has become a central benchmark for quantum algorithms and hardware. The original formulation was developed for qubits, and qubit-based implementations have been demonstrated on several physical platforms~\cite{LONG2001_Grover_NMR, Figgatt2017Grover_trapped_ion, Roy2020trimon, Zhang2022Grover_expt}.

At the same time, many emerging quantum platforms naturally support multilevel quantum systems, or qudits, as elementary information carriers~\cite{nguyen2024empowering, kim2025ultracoherent, Roy2024Qudit, Wang2025transmon12, Low2026qudit_trapped_ion}. Qudits offer a larger local Hilbert space than qubits, which can increase the information density per physical device, reduce the number of carriers required to represent a given database, and, for certain algorithms, reduce the number of entangling operations~\cite{Lanyon2009qudit, Gedik2015qudit, Nikolaeva2021qudit, Nikolaeva2023Toffoli, godwood2026resource}. For search algorithms, this additional structure can also reduce circuit depth and make certain operations more compatible with the native control capabilities of the hardware~\cite{wang2020qudits, Low2026qudit_trapped_ion}. These advantages have motivated both theoretical studies of qudit Grover search~\cite{ivanov2012qudit_Grover, wang2020qudits} and recent experimental realizations using one or two qudits~\cite{Godfrin2017qudit1_Grover, perez2018qudit1_Grover, Roy2023two_qutrit, mohit2025qudit_Grover_expt, Chuang2026qudit_Grover_expt}.

Despite this progress, a unified and hardware-oriented treatment of Grover search for general qudit registers is still needed because realistic devices may have Hilbert-space dimensions that are not powers of two and may combine qudits of different dimensions in heterogeneous architectures~\cite{dogra2015qubit_qutrit_nmr, Litteken2023quantum_waltz, Meth2025lattice_gauge}. Addressing this gap, we formulate Grover search directly in the two-dimensional subspace spanned by the collective target state and its orthogonal complement, allowing the underlying register to consist of qudits with arbitrary dimensions. We review the standard qudit search protocol, connect the abstract oracle and diffusion operators to explicit circuits based on qudit Fourier transforms and controlled-phase gates, and discuss alternative implementations, including ancilla-assisted constructions and diffusion operators that use different conditional phases. We further analyze deterministic phase-matched protocols that achieve unit success probability when the target fraction is known, as well as fixed-point search strategies that avoid overshooting when this information is unavailable. Together, these results provide a flexible framework for implementing quantum search on emerging homogeneous and heterogeneous qudit processors.

\section{Standard Grover search with qudits}

In this section, we first reduce Grover search on an arbitrary qudit register to an effective two-dimensional problem spanned by collective marked and unmarked states. We then derive the standard oracle, diffusion operator, success probability, and optimal iteration count before translating these operations into explicit qudit circuits.

Consider an unsorted database of size $N$ with $M\ge1$ marked entries that satisfy the search criterion. The goal is to find one of these marked entries with the smallest possible number of oracle queries. Unlike a conventional qubit register, a qudit register can naturally encode databases whose sizes are not powers of two. More generally, one may use a heterogeneous register with qudit dimensions $\{d_1,d_2,\ldots\}$, provided the total Hilbert-space dimension $D=\prod_j d_j$ is at least as large as $N$.

\begin{figure*}
    \centering
    \includegraphics[width=\textwidth]{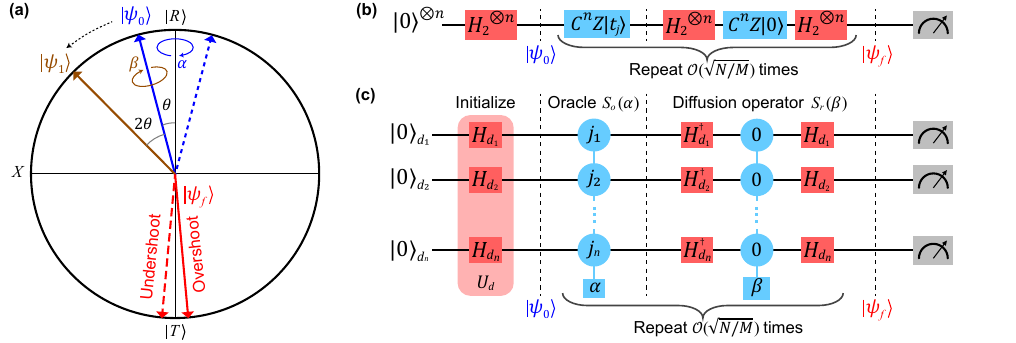}
    \caption{\textbf{Generic Grover search with qudits.} (a) Effective two-level description on the Bloch sphere spanned by the collective unmarked state $\ket{R}$ and target state $\ket{T}$. The oracle $S_o(\alpha)$ applies a phase to the target subspace, while the diffusion operator $S_r(\beta)$ rotates the state about the initial-state axis $\ket{\psi_0}$. For the standard choice $\alpha=\beta=\pi$, the trajectory remains in the $ZX$ plane and generally ends near, but not exactly at, $\ket{T}$. (b) Standard circuit structure with $n$ qubits: prepare $\ket{\psi_0}$, apply $k_{\rm opt}$ oracle--diffusion iterations, and measure the register. (c) Heterogeneous qudit implementation in which the blue layer denotes a generic multi-qudit controlled-phase operation with an arbitrary phase angle, and the red boxes represent qudit Hadamard gates $H_d$.}
    \label{fig:fig1}
\end{figure*}

For the search dynamics, the microscopic encoding can be compressed into two orthogonal subspaces: the marked subspace spanned by the target states $\ket{t_j}$ and the unmarked subspace spanned by the remaining states $\ket{r_j}$. We define the normalized collective states
\begin{subequations}
	\begin{align}
	\ket{T} &= \dfrac{1}{\sqrt{M}} \sum_{j=1}^{M}\ket{t_j}, \\
	\ket{R} &= \dfrac{1}{\sqrt{N-M}} \sum_{j=1}^{N-M}\ket{r_j}.
	\end{align}
\end{subequations}
These orthogonal unit vectors reduce the original search problem to a two-dimensional problem~\cite{brassard2002amp_amplification, Roy2022D2p}. We use the mapping $\ket{R}=\icol{1\\0}$ and $\ket{T}=\icol{0\\1}$, with $2\times2$ matrices denoting the action of the full operators restricted to $\operatorname{span}\{\ket{R},\ket{T}\}$. The search therefore begins in the state
\begin{eqnarray}
\label{eq:psi_0}
	\ket{\psi_0}=\sqrt{1-\lambda} \ket{R} + \sqrt{\lambda} \ket{T} = 
	\begin{bmatrix}
		\sqrt{1-\lambda} \\
		\sqrt{\lambda}
	\end{bmatrix},
\end{eqnarray}
where $\lambda$ is the initial probability of being in the target subspace. For the usual equal-superposition initialization, $\ket{\psi_0}=\frac{1}{\sqrt{N}}\left[ \sum_{j=1}^M\ket{t_j} + \sum_{j=1}^{N-M}\ket{r_j}  \right]$ and therefore $\lambda=M/N$; more generally, the only strict requirement is $\lambda>0$. One convenient unitary that prepares this state from $\ket{R}$ takes the following form in the $\{\ket{R},\ket{T}\}$ basis:
\begin{equation}
U = \begin{bmatrix}
    \sqrt{1-\lambda} & -\sqrt{\lambda} \\
    \sqrt{\lambda} & \sqrt{1-\lambda}
\end{bmatrix}.
\end{equation}
It is convenient to visualize this two-dimensional state as a unit vector on the Bloch sphere, where the north and south poles represent $\ket{R}$ and $\ket{T}$, respectively. Comparing it with the general superposition $\cos(\theta/2)\ket{R}+e^{i\phi}\sin(\theta/2)\ket{T}$, we find that the initial state $\ket{\psi_0}$ makes an angle $\theta=2\sin^{-1}(\sqrt{\lambda})$ with the $z$ axis and lies in the $ZX$ plane, as shown in Fig.~\ref{fig:fig1}(a).

Grover search consists of two basic operations: an oracle that applies a phase to the target subspace and a diffusion, or reflection, operator that amplifies the target amplitude. Each application of the oracle counts as one query, and the scaling of the number of oracle applications with $\lambda$ defines the query complexity of the algorithm. The generalized oracle operator is
\begin{equation}
    S_o(\alpha)=\mathbb{I} - (1-e^{i\alpha})\ket{T}\bra{T} = 
    \begin{bmatrix}
        1 & 0\\
        0 & e^{i\alpha}
    \end{bmatrix},
\end{equation}
where $\mathbb{I}$ is the identity operator. The action of $S_o(\alpha)$ can be visualized as a clockwise rotation of the state vector by an angle $\alpha$ about the $\ket{T}$ axis (or, equivalently, as a counterclockwise rotation about the north-pole axis). The special case $S_o(\alpha=\pi)$ can be viewed as a reflection about the $z$ axis.

The generalized reflection (diffusion) operator about the initial state is
\begin{equation}
\begin{split}
S_r(\beta) &=  \mathbb{I} - (1 - e^{i\beta})\ket{\psi_0}\bra{\psi_0} \\
&=
\begin{bmatrix}
    1 & 0 \\
    0 & 1
\end{bmatrix} - 
(1 - e^{i\beta})
\begin{bmatrix}
\sqrt{1-\lambda} & \sqrt{\lambda}
\end{bmatrix}
\begin{bmatrix}
    \sqrt{1-\lambda} \\
		\sqrt{\lambda}
\end{bmatrix} \\
&=
\begin{bmatrix}
\lambda +e^{i\beta}(1-\lambda) & (e^{i\beta}-1) \sqrt{\lambda (1-\lambda)} \\
(e^{i\beta}-1) \sqrt{\lambda (1-\lambda)} & e^{i\beta}\lambda + (1-\lambda)
\end{bmatrix},
\end{split}
\label{eq:reflect}
\end{equation}
which can be visualized as a clockwise rotation of the current state vector about the $\ket{\psi_0}$ axis by an angle $\beta$.

The product of the oracle and reflection operators is often called the Grover iterate, $G(\alpha,\beta)=S_r(\beta)S_o(\alpha)$. We omit the negative global phase commonly used in the literature~\cite{Chuang2014fixed_point, Li2007phase_matching} because it has no physical effect. The standard Grover iterate with $\alpha=\pm \pi$ and $\beta=\pm \pi$ rotates the state vector by $2\theta$ about the $y$ axis, as shown in Fig.~\ref{fig:fig1}(a), restricting the trajectory to the $ZX$ plane. Consequently, the rotated state vector becomes $\ket{\psi_1}=\cos(3\theta/2)\ket{R}+\sin(3\theta/2)\ket{T}$, resulting in a success probability of $|\langle T|\psi_1\rangle|^2=\sin^2(3\theta/2)$. In general, the success probability after the $k$th iteration is
\begin{equation}
    P_{\rm succ}(k)=|\langle T|\psi_k\rangle|^2=\sin^2\left[\left(k+\tfrac{1}{2}\right)\theta\right],
\end{equation}
where 
\begin{equation}
    \ket{\psi_k} = \cos{\left[\left(k+\tfrac{1}{2}\right)\theta\right]}\ket{R}+\sin{\left[\left(k+\tfrac{1}{2}\right)\theta\right]}\ket{T}.
\label{eq:psi_k}
\end{equation}
Clearly, the success probability is periodic in the number of iterations.
Since the angular distance of $\ket{\psi_0}$ from the south pole is $\pi-\theta$, the number of steps needed to reach $\ket{T}$ is $(\pi-\theta)/(2\theta)$, which is generally not an integer. Consequently, the final state $\ket{\psi_f}$ will ``undershoot'' or ``overshoot'' the target state if the fractional part of this ratio is less than or greater than $0.5$, respectively. Therefore, the optimal number of steps for the original Grover search is the nearest integer~\cite{Long2001certain_Grover}
\begin{equation}
k_{\rm opt} = \left \lfloor \dfrac{\pi}{2\theta} - \dfrac{1}{2} \right \rceil = \left \lfloor \dfrac{\pi}{4\sin^{-1}\sqrt{\lambda}} - \dfrac{1}{2} \right \rceil = \left \lfloor \dfrac{\pi}{4\sin^{-1}\sqrt{\lambda}} \right \rfloor,
\end{equation}
where $\lfloor \boldsymbol\cdot \rceil$ denotes rounding to the nearest integer and $\lfloor \boldsymbol\cdot \rfloor$ denotes the floor function.

After the optimal number of iterations, the system is measured. If there is only one target state ($M=1$), the system collapses to that state with probability $P_{\rm succ}(k_{\rm opt})$. For multiple correct answers, each marked state is obtained with probability $P_{\rm succ}(k_{\rm opt})/M$. For example, when $\lambda<0.002$, the success probability $P_{\rm succ}(k_{\rm opt})$ exceeds 0.99 for the standard Grover search. For small $\lambda$ (corresponding to a large database),
\begin{equation}
    k_{\rm opt}\approx\frac{\pi}{4}\sqrt{\frac{N}{M}},
\end{equation}
which gives the quadratic speedup compared with classical search.

\subsection{Implementation without an ancilla}
The quantum circuit for generalized qudit Grover search is shown in Fig.~\ref{fig:fig1}(c) and can be compared with the standard qubit circuit in Fig.~\ref{fig:fig1}(b). It involves three stages: prepare a superposition state with nonzero overlap with the answer, apply the oracle--diffusion pair $k_{\rm opt}$ times, and measure the register. A common way to prepare the equal-superposition state is to apply the qudit Hadamard, or discrete Fourier-transform, gate
\begin{equation}
H_d \;=\; \frac{1}{\sqrt{d}}
\sum_{j=0}^{d-1} \sum_{k=0}^{d-1}
\omega^{jk} \, |j\rangle \langle k|,
\qquad
\omega = e^{2\pi i / d},
\end{equation}
to the ground state $\ket{0_d}$. We use an explicit subscript $d$ to indicate states or gates in the $d$-dimensional Hilbert space. For a heterogeneous system consisting of qudits with dimensions $\{d_1,d_2,\ldots,d_n\}$, one can construct the unitary $U_d=H_{d_1}\otimes H_{d_2}\otimes\cdots\otimes H_{d_n}$ by applying the individual Hadamard gates. The requirements are that $U_d$ be unitary and that $\braket{T}{U_d}{0_d}\neq0$. Here, we assume $N\le D=\prod_j d_j$ and treat all unused $D-N$ states as additional unmarked items included in $\ket{R}$, so that $\lambda=M/D$. The iteration count depends only on $\lambda$; the qudit structure affects the implementation cost per iteration, but not the number of queries.

The oracle $S_o(\pm\pi)$ flips the phase of the target state, so that $\ket{T}\mapsto-\ket{T}$. It can be realized by applying one or more qudit conditional-phase gates (if $M>1$), each of which flips the phase of a target state $\ket{m}$ while leaving the other basis states unchanged: $\ket{m}\mapsto-\ket{m}$ and $\ket{k}\mapsto\ket{k}$ for $k\neq m$. Such a gate is defined as
\begin{equation}
    \mathtt{CPHASE}_{d,m} = \mathbb{I}_d - 2\ket{m}\bra{m},
\end{equation}
where $\mathbb{I}_d$ is the $d$-dimensional identity matrix and $\ket{m} \equiv\bigotimes_j \ket{m_j}$. This $\mathtt{CPHASE}$ gate should not be confused with the conditional-$Z$, or $\mathtt{CZ}$, gate for qudits (see Appendix~\ref{app:CZ}).

The standard Grover diffusion operator, which can be viewed as a reflection of the current state about the axis $\ket{\psi_0}$, can be decomposed using the other qudit operations discussed above:
\begin{equation}
\begin{split}
    S_r(\pi) &= \mathbb{I}_d - 2\ket{\psi_0}\bra{\psi_0} \\
    &= U_d \cdot U_d^\dagger - 2U_d \cdot \ket{0_d}\bra{0_d} \cdot U_d^\dagger \\
    &= U_d \cdot (\mathbb{I}_d - 2\ket{0_d}\bra{0_d} ) \cdot U_d^\dagger \\
    &= U_d \cdot \mathtt{CPHASE}_{d,0} \cdot U_d^\dagger.
\end{split}
\end{equation}
The last equation shows that the reflection operator can be constructed using the unitary that prepares the initial superposition state and a controlled-phase gate that acts only on the ground state $\ket{0_d}$. Note the requirement of using $U_d^\dagger$ since $U_d^\dagger \neq U_d$ in general, unlike the qubit case where $H_2^\dagger = H_2$.

\subsection{Use of alternative conditional-phase gates}
The phase gate in the diffusion operator need not act on $\ket{0}$. Instead, one may choose a gate that acts on any basis state $\ket{m}$ convenient for an experimental implementation (see Appendix~\ref{app:diffusion}). One can show that
\begin{equation}
\begin{split}
    S_r(\pi) &= H_d \cdot (\mathbb{I}_d - 2\ket{0_d}\bra{0_d} ) \cdot H_d^\dagger \\
    &= H_d \cdot X_d^{d-m} \cdot (\mathbb{I}_d - 2\ket{m}\bra{m} ) \cdot (X_d^{\dagger})^{d-m} \cdot H_d^\dagger \\
    &= Z_d^{d-m} \cdot H_d \cdot (\mathbb{I}_d - 2\ket{m}\bra{m} ) \cdot H_d^\dagger \cdot (Z_d^{\dagger})^{d-m} \\
    &= Z_d^{d-m} \cdot H_d \cdot \mathtt{CPHASE}_{d,m} \cdot H_d^\dagger \cdot (Z_d^{\dagger})^{d-m} \\
    &= (Z_d^{\dagger})^{m} \cdot H_d \cdot \mathtt{CPHASE}_{d,m} \cdot H_d^\dagger \cdot (Z_d)^{m}
\end{split}
\end{equation}
where the qudit phase and cyclic shift gates (generalized $\mathtt{NOT}$) are defined as
\begin{align}
    Z_d &= \sum_{k=0}^{d-1}\omega^{k} \, \ket{k}\bra{k}, \\
    X_d &= \sum_{k=0}^{d-1}\ket{k+1 \bmod d}\bra{k},
\end{align}
and we have used the identities $H_d X_d^m=Z_d^m H_d$ and $(Z_d^{\dagger})^{d-m}=Z_d^{m}$ (see Appendix~\ref{app:identities} for proofs). The additional single-qudit $Z$ gates can increase the circuit depth by at most two. On many platforms, such as superconducting systems, the $Z$ gates are virtual~\cite{Litteken2023quantum_waltz, Wang2025transmon12} and do not affect the algorithm runtime.

\subsection{Implementation with an ancilla}
In certain realizations, it may be convenient to use an ancilla to implement the qudit conditional-phase gate. The general idea is to obtain phase kickback by preparing the ancilla in the negative eigenstate of a chosen unitary. Even for qudit Grover search, only a qubit ancilla is needed, independent of the database size. We initialize the ancilla in the $(\ket{0}-\ket{1})/\sqrt{2}$ superposition state by applying $R_y(-\pi/2)$ or $H_2\cdot X_2$ to $\ket{0}$. The effective phase gate $\mathtt{CPHASE}_{d,m}$ is implemented by applying a conditional-NOT gate $\mathtt{CX}_{d,m}$ to the ancilla qubit when the control qudit is in state $\ket{m}$:
\begin{equation}
    \mathtt{CX}_{d,m} = \sum_{k\neq m}|k\rangle\langle k|\otimes\mathbb{I}_2+|m\rangle\langle m|\otimes X_2.
\end{equation}
Writing the unnormalized ancilla state as $\ket{-}=\ket{0}-\ket{1}$ and the data state as $\ket{\psi_d}=\sum_k c_k\ket{k}$, we obtain
\begin{equation}
\begin{split}
    \mathtt{CX}_{d,m}\left[\ket{\psi_d} \otimes\ket{-}\right]
    =& \; \mathtt{CX}_{d,m}\left( \sum_k c_k\ket{k} \otimes \ket{-} \right) \\
    =& \; \sum_{k\neq m}c_k\ket{k}\otimes\ket{-} - c_m\ket{m}\otimes\ket{-} \\
    =& \; (\mathbb{I}_d - 2\ket{m}\bra{m}) \sum_k c_k\ket{k}\otimes\ket{-} \\
    =& \; \left[ \mathtt{CPHASE}_{d,m}\ket{\psi_d} \right] \otimes \ket{-},
\end{split}
\end{equation}
where the normalization factor of the ancilla state has been omitted for brevity.

\section{Deterministic Grover search}

As discussed in the previous section, the standard Grover protocol generally stops with a success probability $P_{\rm succ}(k_{\rm opt})$ that is close to, but not exactly, unity. This residual error arises because the ideal rotation angle needed to align the Bloch vector with $\ket{T}$ is usually not an integer multiple of the standard Grover rotation $2\theta$. A natural way to remove this undershoot or overshoot is to adjust the phases in the oracle and/or diffusion operators, a technique commonly known as phase matching~\cite{Long1999phase_matching, Li2007phase_matching, toyama2008multiphase}.

\begin{figure*}
    \centering
    \includegraphics[width=\textwidth]{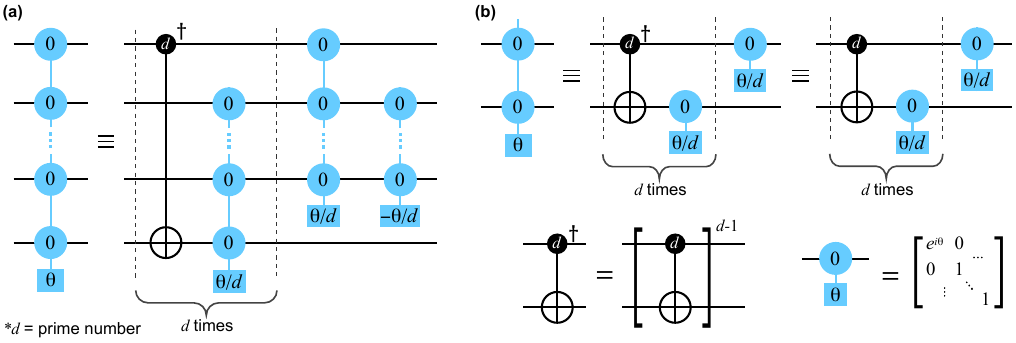}
    \caption{\textbf{Decomposition of a multi-controlled qudit $\mathtt{CPHASE}(\theta)$ gate for qudits with identical prime dimension $d$.} (a) The decomposition involves two-qudit $\mathtt{CSUM}^{\dagger}$ gates and $\mathtt{CPHASE}(\theta)$ gates with fewer controls. (b) Decomposition of the two-qudit $\mathtt{CPHASE}(\theta)$ gate (top). The bottom panel shows the identity $\mathtt{CSUM}^{\dagger}=\mathtt{CSUM}^{d-1}$ and the single-qudit phase gate acting on basis state $\ket{0}$.}
    \label{fig:cphase}
\end{figure*}

For exact search, it is useful to define
\begin{equation}
k_{\rm certain} = \left \lceil \dfrac{\pi}{2\theta} - \dfrac{1}{2} \right \rceil = \left \lceil \dfrac{\pi}{4\sin^{-1}\sqrt{\lambda}} - \dfrac{1}{2} \right \rceil,
\end{equation}
where $\lceil \boldsymbol\cdot \rceil$ denotes the ceiling function. One can verify that $k_{\rm certain}-k_{\rm opt}$ is either 0 or 1; the latter case corresponds to an undershoot, where the standard protocol stops just before reaching the south pole of the Bloch sphere. In this section, we describe four deterministic strategies for reaching $\ket{T}$ exactly. The first three can be implemented without an ancilla, while the fourth uses a single qubit ancilla to tune the effective initial target overlap. In all cases, the qudit phase inversion $\mathtt{CPHASE}_{d,m}$ is generalized to a phase rotation
\begin{equation}
    \mathtt{CPHASE}_{d,m}(\phi) = \mathbb{I}_d - (1-e^{i\phi})\ket{m}\bra{m}.
\end{equation}
Note that the $\mathtt{CPHASE}_{d,m}(\phi)$ gate can be implemented as a native operation on several bosonic quantum-computing platforms, including superconducting cavity-resonator and trapped-ion systems, using selective number-dependent arbitrary phase (SNAP) gates~\cite{Park2024qsim, SNAP2015PRL, Bornman2025benchmark}. Consequently, employing this gate directly can significantly reduce the circuit depth compared with recursive decompositions into elementary two-qudit and single-qudit gates, such as that shown in Fig.~\ref{fig:cphase}. Here, we assume that all $n$ qudits have the same prime dimension $d$, with the two-qudit $\mathtt{CSUM}$ gate defined as
\begin{equation}
    \mathtt{CSUM} = \sum_{j=0}^{d-1}\ket{j}\bra{j} \otimes X_{d}^j.
\end{equation}
This decomposition requires $\mathcal{O}((d+1)^n)$ two-qudit gates without using any ancilla (see Appendix~\ref{app:recursion}).

\subsection{Method 1}
\begin{figure*}
    \centering
    \includegraphics[width=\textwidth]{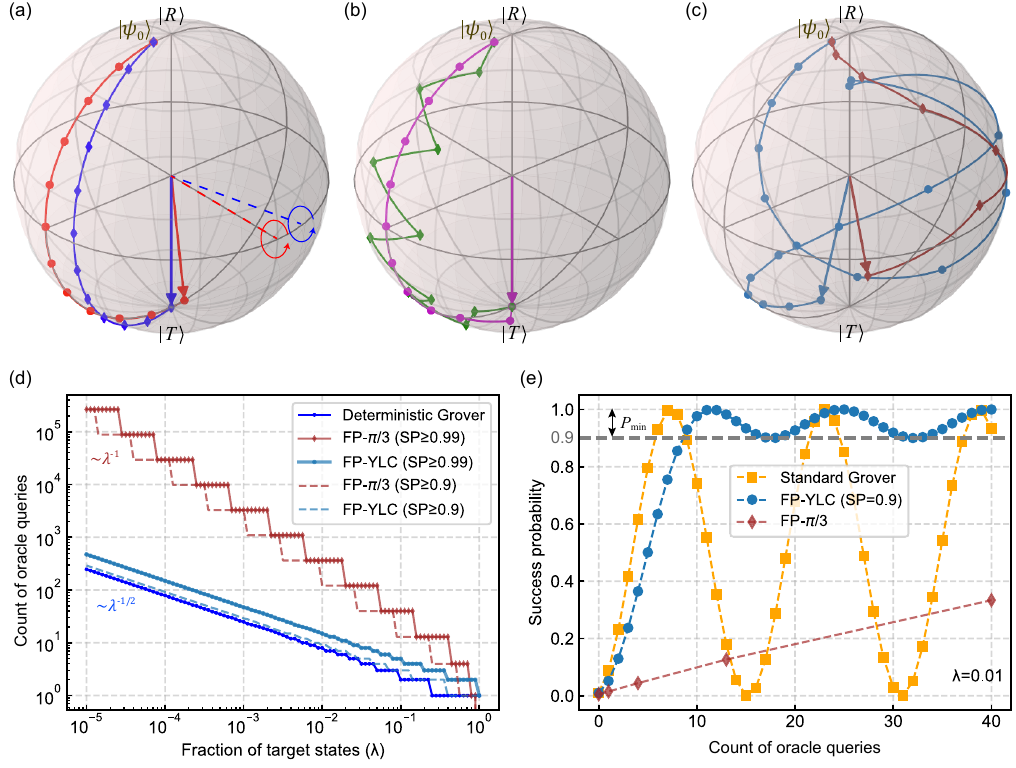}
    \caption{\textbf{Comparison of quantum search protocols.} Bloch-sphere trajectories illustrate how different phase choices steer the state toward the target state: (a) standard Grover search (red circles) and the one-parameter deterministic protocol D1p (blue diamonds), (b) the two-parameter D2p (green diamonds) and three-parameter D3p (magenta circles) deterministic protocols, and (c) the $\pi/3$ (brown diamonds) and YLC (blue circles) fixed-point (FP) protocols. The initial state $\ket{\psi_0}$ is prepared with $\lambda=0.005$. (d) Oracle-query counts versus target-state fraction $\lambda$ for deterministic and fixed-point searches. Solid and dashed fixed-point curves correspond to success-probability thresholds of 0.99 and 0.9, respectively. The query complexities of the YLC and $\pi/3$ protocols scale as $\mathcal{O}(1/\sqrt{\lambda})$ and $\mathcal{O}(1/\lambda)$, respectively, as illustrated by the light-blue and brown traces. (e) Success probability versus the number of oracle queries for standard Grover (orange squares), FP-YLC (blue circles), and FP-$\pi/3$ (brown diamonds) searches. FP-YLC remains above the chosen threshold $P_{\rm min}$ after $k_{\rm min}$ iterations, whereas the FP-$\pi/3$ protocol is defined only at query counts $(3^k-1)/2$ because of its recursive construction. The parameters are $\lambda=0.01$ and $P_{\rm min}=0.9$, yielding $k_{\rm min}=9$.}
    \label{fig:fig3}
\end{figure*}

The first method, introduced by Long et al.~\cite{Long2001certain_Grover}, modifies both the oracle and diffusion phases by using the same angle in every Grover iterate. Specifically, one applies $G(\theta_c,\theta_c)$ with
\begin{equation}
\label{eq:d1p_theta}
    \theta_c = 2 \sin^{-1} \left( \frac{1}{\sqrt{\lambda}} \sin{\left(\frac{\pi}{4k_{\rm certain}+2} \right)} \right),
\end{equation}
so that
\begin{equation}
    \ket{T}=\left[ G(\theta_c, \theta_c) \right]^{k_{\rm certain}} \ket{\psi_0}.
\end{equation}
Replacing the standard phase inversion by the generalized controlled-phase gate $\mathtt{CPHASE}_{d,m}(\theta_c)$ causes the Bloch vector to follow an arc outside the $ZX$ plane and land exactly on the south pole after an integer number of steps, as shown by the blue trace in Fig.~\ref{fig:fig3}(a). The construction also works for any number of steps larger than $k_{\rm certain}$, provided $\theta_c$ is chosen accordingly. Its main experimental advantage is that the same phase parameter $\theta_c$ is used in every iteration, unlike in more general multiparameter phase-matching schemes~\cite{Braunstein2007exact, Toyama2013certain_Grover}. We refer to this protocol as deterministic one-parameter Grover search, or D1p.

\subsection{Method 2}
The D1p protocol requires adjustment of the oracle phase $\alpha$, which may not be possible when the oracle is fixed by the problem or hardware implementation. To address this constraint, Roy et al. developed a deterministic two-parameter protocol (D2p) that leaves the oracle $S_o(\alpha)$ unchanged and adjusts the diffusion operator to achieve exact search~\cite{Roy2022D2p}. The scheme works for any $\alpha\neq0$ and is most efficient for $\alpha\simeq\pi$, reaching certainty in $k_{\rm certain}$ steps when $\alpha=\pm\pi$.

For the standard oracle phase $\alpha=\pm\pi$, D2p alternates between two diffusion phases, $\beta_1$ and $\beta_2$, determined numerically for a given $\lambda$. The sequence is
\begin{equation}
    \ket{T} = \underbrace{\cdots G(\pi,\beta_1)\cdot G(\pi,\beta_2)\cdot G(\pi,\beta_1)}_{k_{\rm certain}\; \text{times}} \ket{\psi_0},
\end{equation}
which produces a zigzag trajectory around the geodesic in the $ZX$ plane. The green trace in Fig.~\ref{fig:fig3}(b) shows an example with $\lambda=0.005$, requiring $\beta_1=2.616$, $\beta_2=-2.606$, and $k_{\rm certain}=11$.

When $k_{\rm even}\ge k_{\rm certain}$ is even, the phase values satisfy
\begin{subequations}
	\begin{align}
	1 + 4\lambda (1-2\lambda) \sin\left( \dfrac{\beta_1}{2} \right) \sin\left( \dfrac{\beta_2}{2} \right) \dfrac{\tan(\frac{k_{\rm even}}{2} \phi)}{\sin(\phi)} &= 0, \\
	(1-4\lambda) \tan\left( \dfrac{\beta_1}{2} \right) + \tan\left( \dfrac{\beta_2}{2} \right) &= 0,
	\end{align}
\label{eq:thetas}
\end{subequations}
where
\begin{equation}
	\cos(\phi) = \cos\left( \dfrac{\beta_1+\beta_2}{2} \right) + 8\lambda(1-\lambda) \sin\left( \dfrac{\beta_1}{2} \right) \sin\left( \dfrac{\beta_2}{2} \right).
\end{equation}
The corresponding equations for an odd number of iterations~\cite{Roy2023D2p_erratum} are given in Appendix~\ref{app:equations}.

A useful dual version keeps the diffusion phase fixed at $\pm\pi$ and alternates the oracle phase between the same two values,
\begin{equation}
    \ket{T} = \underbrace{G(\beta_1, \pi)\cdot G(\beta_2, \pi)\cdot G(\beta_1, \pi) \cdots }_{k_{\rm certain}\; \text{times}} \ket{\psi_0}.
\end{equation}
Note the order of the $\beta$ parameters here. The trajectory again follows a zigzag path before reaching the south pole, and the alternating structure may provide experimental advantages through partial noise cancellation~\cite{mohit2025qudit_Grover_expt}.

\subsection{Method 3}
The D2p idea can be further simplified by using standard Grover iterations for most of the protocol and modifying only the final steps. Here, we introduce a deterministic three-parameter protocol (D3p). The first $k'$ steps use the standard diffusion operator $S_r(\pi)$, while the remaining $k_{\rm certain}-k'\ge2$ steps alternate between two phases $\beta_a$ and $\beta_b$:
\begin{equation}
    \ket{T} = \underbrace{G(\pi,\beta_a)\cdot G(\pi,\beta_b)\cdot G(\pi,\beta_a) \cdots}_{k_{\rm certain} - k'\; \text{times}} \; [G(\pi,\pi)]^{k'} \ket{\psi_0}.
\label{eq:d3p}
\end{equation}
The intermediate state $[G(\pi,\pi)]^{k'}\ket{\psi_0}=\ket{\psi_{k'}}$ follows directly from Eq.~\eqref{eq:psi_k}, and the remaining phase parameters are determined numerically. The magenta trace in Fig.~\ref{fig:fig3}(b) shows an example in which only the final two steps are modified and the state reaches the target with certainty (see Appendix~\ref{app:equations} for the corresponding equations).

\subsection{Method 4}
The fourth method, developed for qubits by Mishra et al.~\cite{Mishra2026deterministic}, uses an ancilla to tune the effective target overlap. The same idea extends naturally to qudit registers, and the ancilla only needs to be a two-level system. The goal is to choose an effective overlap $\lambda'\le\lambda$ such that
\begin{equation}
k_{\rm certain} =  \dfrac{\pi}{4\sin^{-1}\sqrt{\lambda'}} - \dfrac{1}{2},
\end{equation}
so that the exact-search condition is satisfied.

To realize this smaller overlap, initialize the ancilla in $a\ket{0}+b\ket{1}$ with $|a|^2+|b|^2=1$ and choose $|a|^2\lambda=\lambda'$. Equivalently, one can apply a rotation $R_y(\theta_a)$ with
\begin{equation}
    \theta_a=2 \cos^{-1}\left( \sqrt{\frac{\lambda'}{\lambda}} \right),
\end{equation}
which gives $|a|^2=[\cos(\theta_a/2)]^2=\lambda'/\lambda$. The joint initial state is then
\begin{equation}
\begin{split}
    \ket{\psi_0'} =& \left[ \sqrt{1-\lambda} \ket{R} + \sqrt{\lambda} \ket{T} \right] \otimes \left[ a\ket{0}_2 + \sqrt{1-|a|^2}\ket{1}_2 \right]  \\
    =& \; U_d \cdot \ket{0}_d \otimes R_y(\theta_a)\cdot \ket{0}_2 \\
    =& \; U_d \otimes R_y(\theta_a) \cdot \ket{0}_d \otimes \ket{0}_2 \\
    =& \; U_d' \cdot \ket{0}_d \otimes \ket{0}_2.
\end{split}
\end{equation}
The target final state is $\ket{T}\otimes\ket{0}_2$. Accordingly, the $\mathtt{CPHASE}$ gate is modified so that it is controlled on the ancilla state $\ket{0}_2$, and the diffusion operator is constructed using $U_d'^\dagger=U_d^\dagger \otimes R_y(-\theta_a)$. One may also choose any other rotation axis in the $XY$ plane for the qubit rotation.

It is similarly possible to use $\ket{1}_2$ as the ancilla control state by choosing $|b|^2\lambda=[\sin(\theta_a/2)]^2\lambda=\lambda'$. In that case, the final state is $\ket{T}\otimes\ket{1}_2$. This freedom allows the implementation to be adapted to whichever controlled-phase operation has the higher fidelity on a given platform.

\section{Fixed-point search}

The deterministic protocols above assume that the initial target fraction $\lambda$ is known. This information is needed to choose the number of Grover iterates and, when necessary, the phase parameters that prevent the Bloch vector from undershooting or overshooting $\ket{T}$. In many search problems, however, $\lambda$ may not be known \textit{a priori}. In that setting, it is useful to design Grover-like iterates that either avoid overshooting altogether or guarantee that the success probability remains above a chosen threshold $P_{\rm min}$ once the number of iterations is sufficiently large. Such protocols are known as fixed-point search algorithms.

\subsection{The \texorpdfstring{$\pi/3$}{pi/3} algorithm}
Grover's fixed-point protocol~\cite{Grover2005piby3} replaces the usual phase inversions in both the oracle and diffusion operators by smaller rotations with $\alpha=\beta=\pi/3$. The resulting transformation is constructed recursively as
\begin{equation}
    U_{k+1} = U_k S_r(\pi/3)U_k^\dagger S_o(\pi/3)U_k,
\end{equation}
where $U_k$ acts on $\ket{\psi_0}$. The initial operator $U_0$ is chosen such that $|\braket{T}{U_0}{\psi_0}|^2=\lambda>0$; it may, for example, be $\mathbb{I}$. Because each recursion amplifies the target component without rotating past it, the state approaches $\ket{T}$ monotonically for any nonzero $\lambda$. The corresponding success probability is
\begin{equation}
    P_{\rm succ}(k)=1-(1-\lambda)^{3^k}.
\end{equation}

This monotonic behavior comes at the cost of query complexity. Unlike the standard and deterministic protocols, for which the number of oracle calls equals the iteration count, the recursive construction requires $(3^k-1)/2$ oracle calls after $k$ recursions. Consequently, the scaling becomes $\mathcal{O}(1/\lambda)$ rather than the quadratic $\mathcal{O}(1/\sqrt{\lambda})$ scaling of Grover search. The numbers of oracle calls needed to reach target success probabilities of 0.9 and 0.99 are shown by the brown traces in Fig.~\ref{fig:fig3}(d), and the corresponding Bloch-sphere evolution for $\lambda=0.005$ through $k=6$ is shown in Fig.~\ref{fig:fig3}(c).

\subsection{The YLC algorithm}
The Yoder--Low--Chuang (YLC) algorithm~\cite{Chuang2014fixed_point} preserves the quadratic speedup while retaining the robustness of fixed-point search. Instead of using the same phase at every step, the YLC protocol chooses a different pair of phases for each Grover iterate so that the success probability is bounded below by a user-chosen value $P_{\rm min}$ for all $\lambda>\lambda_{\rm min}$ once the iteration count exceeds a threshold $k_{\rm min}$. Defining $\delta=\sqrt{1-P_{\rm min}}$, this threshold is approximately
\begin{equation}
    k_{\rm min} \approx \Big\lceil \frac{1}{2\sqrt{\lambda_{\rm min}}}\log\left(\frac{2}{\delta}\right) -\frac{1}{2}\Big\rceil,
\label{eq:chuang_kmin}
\end{equation}
which retains the characteristic quadratic scaling.

For a fixed iteration count $k$, the YLC phases are chosen in a palindromic order. For $j\in\{1,2,\ldots,k\}$,
\begin{equation}
    \alpha_j = \beta_{k-j+1} = 2 \cot^{-1}\left( \sqrt{1-\gamma^2} \tan \left(\frac{2\pi j}{2k+1} \right) \right),
\end{equation}
where $\gamma^{-1}=T_{1/(2k+1)}(1/\delta)$ and $T_n(x)$ is the $n$-th Chebyshev polynomial of the first kind (see Appendix~\ref{app:cheby}). Equivalently, for a desired $P_{\rm min}$ and a chosen iteration count $k$, the smallest target fraction for which the bound is guaranteed is
\begin{equation}
    \lambda_{\rm min} = 1 - \left( T_{1/(2k+1)}(1/\delta) \right)^{-2} \approx \left[ \frac{1}{2k+1}\log\left(\frac{2}{\delta}\right) \right]^2.
\end{equation}

As shown by the blue symbols in Fig.~\ref{fig:fig3}(e), the success probability then oscillates between $P_{\rm min}$ and 1 for $k>k_{\rm min}$ rather than falling below the threshold because of overshooting. In our notation, the oracle count is half of that used in Ref.~\cite{Chuang2014fixed_point}, where the generalized oracle $S_o(\alpha)$ was decomposed using an ancilla qubit and therefore required two applications of the standard oracle $S_o(\pi)$. The blue trace in Fig.~\ref{fig:fig3}(c) shows the Bloch-sphere trajectory for $\lambda=0.005$ and $P_{\rm min}=0.98$ through $k=19$.

Figure~\ref{fig:fig3}(d) compares the oracle counts for deterministic search and the two fixed-point protocols as a function of $\lambda$. The YLC protocol follows the same $\mathcal{O}(1/\sqrt{\lambda})$ scaling as deterministic Grover search, with an offset set by the chosen threshold $P_{\rm min}$, whereas the simpler $\pi/3$ protocol provides monotonic convergence but scales as $\mathcal{O}(1/\lambda)$.

\section{Conclusion}

Qudits are rapidly moving from a theoretical resource to an experimental platform, with recent demonstrations of universal qudit gate sets and increasingly coherent multilevel devices~\cite{kim2025ultracoherent, Ringbauer2022universal_qudit, huang2026universal_qudit}. In this work, we have developed a unified formulation of Grover search for qudit-based architectures, including heterogeneous registers composed of qudits with different dimensions. Regardless of the microscopic encoding, the search dynamics remain confined to the two-dimensional space spanned by the collective target state $\ket{T}$ and its orthogonal complement $\ket{R}$. This reduction provides a geometric description of the algorithm while connecting the abstract reflections to experimentally relevant qudit gates.

We have shown that the standard Grover iterate can be implemented using qudit Fourier transforms and controlled-phase operations, with equivalent constructions available for different choices of conditional-phase gates. These alternatives are useful because the native high-fidelity entangling operation may vary among hardware platforms. We also described four exact-search strategies: the one-parameter deterministic protocol (D1p) with matched oracle and diffusion phases, the two-parameter protocol (D2p), which can keep the oracle fixed, the three-parameter protocol (D3p), which modifies only the final diffusion steps after standard Grover iterations, and an ancilla-assisted method that tunes the effective target overlap. These methods can identify a correct answer with certainty using at most one additional oracle query compared with the standard protocol.

When the target fraction is known, deterministic phase-matched protocols can remove the residual undershoot or overshoot and reach the target subspace with unit success probability while preserving the quadratic query scaling. When the target fraction is unknown, fixed-point protocols provide robustness by preventing destructive overshooting and guaranteeing a prescribed minimum success probability above a threshold value of $\lambda$. In particular, the YLC construction retains the characteristic $\mathcal{O}(1/\sqrt{\lambda})$ scaling, whereas the simpler $\pi/3$ protocol provides monotonic amplification at the cost of losing the quadratic speedup. We also provided a comparison of the corresponding trajectories on the Bloch sphere for the different methods, offering a useful visualization for understanding their similarities and trade-offs.

These results provide a compact toolbox for designing search algorithms on qudit processors. The choice among standard, deterministic, ancilla-assisted, and fixed-point protocols should be guided by access to generalized phase gates, controlled-gate fidelity, knowledge of the target fraction, and acceptable circuit depth. As qudit hardware matures, Grover search will remain valuable both as a near-term benchmark for multilevel quantum control and as a future application that uses larger local Hilbert spaces to encode and search larger problem instances efficiently. Beyond computation, Grover-like qudit amplitude amplification can also improve quantum sensing by reaching the Grover--Heisenberg limit, a fundamental lower bound on the minimum sensing time~\cite{allen2025GHL}.

\begin{acknowledgments}
This work was supported by the U.S. Department of Energy, Office of Science, National Quantum Information Science Research Centers, Superconducting Quantum Materials and Systems Center (SQMS), under Contract No. 89243024CSC000002. Fermilab is operated by Fermi Forward Discovery Group, LLC under Contract No. 89243024CSC000002 with the U.S. Department of Energy, Office of Science, Office of High Energy Physics.

\end{acknowledgments}


\renewcommand{\thetable}{A\arabic{table}}
\setcounter{table}{0}
\renewcommand{\thefigure}{A\arabic{figure}}
\renewcommand{\theHfigure}{A\arabic{figure}}
\setcounter{figure}{0}
\renewcommand{\theequation}{A\arabic{equation}}
\setcounter{equation}{0}

\appendix
\numberwithin{equation}{section}

\section{Equivalence of \texorpdfstring{$\mathtt{CZ}$ and $\mathtt{CSUM}$}{CZ and CSUM}}
\label{app:CZ}

Consider two qudits that may have unequal dimensions. The generalized conditional-$Z$ gate is defined as
\begin{equation}
    \mathtt{CZ} = \sum_{j=0}^{d_1-1}\ket{j}\bra{j} \otimes Z_{d_2}^j,
\end{equation}
The $\mathtt{CZ}$ and generalized $\mathtt{CSUM}$ gates are related by conjugating the target qudit with qudit Hadamard gates. The $\mathtt{CSUM}$ gate is the qudit generalization of the two-qubit $\mathtt{CNOT}$ gate and is defined as
\begin{equation}
    \mathtt{CSUM} = \sum_{j=0}^{d_1-1}\ket{j}\bra{j} \otimes X_{d_2}^j.
\end{equation}

Their equivalence follows from
\begin{equation}
\begin{split}
    (\mathbb{I}_{d_1} & \otimes H_{d_2}) \cdot  \mathtt{CSUM} \cdot (\mathbb{I}_{d_1} \otimes H_{d_2}^\dagger) \\
   & = (\mathbb{I}_{d_1} \otimes H_{d_2}) \cdot \left( \sum_{j=0}^{d_1-1}\ket{j}\bra{j} \otimes X_{d_2}^j \right) \cdot (\mathbb{I}_{d_1} \otimes H_{d_2}^\dagger) \\
   & = \sum_{j=0}^{d_1-1}\ket{j}\bra{j} \otimes (H_{d_2} \cdot X_{d_2}^j  \cdot H_{d_2}^\dagger ) \\
   & = \sum_{j=0}^{d_1-1}\ket{j}\bra{j} \otimes Z_{d_2}^j = \mathtt{CZ},
\end{split}
\end{equation}
where the last line uses the identity $H_d  X_d  H_d^\dagger = Z_d$.

The $\mathtt{CZ}$ gate should be distinguished from the single-qudit conditional-phase gate $\mathtt{CPHASE}_{d,m}(\theta)$. Below, we explicitly show the shift, Hadamard, and conditional-phase gates for a ququint ($d=5$):
\begin{equation}
    X_5=\begin{bmatrix}
        0 & 0 & 0 & 0 & 1 \\
        1 & 0 & 0 & 0 & 0 \\
        0 & 1 & 0 & 0 & 0 \\
        0 & 0 & 1 & 0 & 0 \\
        0 & 0 & 0 & 1 & 0 \\
    \end{bmatrix}, \qquad
\end{equation}
\begin{equation}
    H_5=\frac{1}{\sqrt{5}}\begin{bmatrix}
        1 & 1 & 1 & 1 & 1\\
        1 & \omega & \omega^2 & \omega^3 & \omega^4 \\
        1 & \omega^2 & \omega^4 & \omega & \omega^3\\
        1 & \omega^3 & \omega & \omega^4 & \omega^{2}\\
        1 & \omega^4 & \omega^3 & \omega^{2} & \omega
    \end{bmatrix}, \; \omega=e^{2\pi i/5}
\end{equation}
\begin{equation}
    \mathtt{CPHASE}_{5,0}(\theta)=\begin{bmatrix}
        e^{i\theta} & 0 & 0 & 0 & 0 \\
        0 & 1 & 0 & 0 & 0 \\
        0 & 0 & 1 & 0 & 0 \\
        0 & 0 & 0 & 1 & 0 \\
        0 & 0 & 0 & 0 & 1 \\
    \end{bmatrix}. \qquad \qquad \; \;
\end{equation}


\section{Alternative diffusion operator}
\label{app:diffusion}
\begin{figure}[H]
    \centering
    \includegraphics[width=\columnwidth]{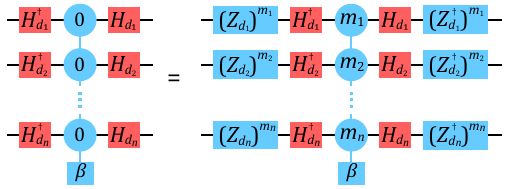}
    \caption{\textbf{Alternative qudit diffusion operator.} Circuit identity for implementing the diffusion reflection with a controlled-phase gate acting on an arbitrary basis state $\ket{m}$ instead of the standard state $\ket{0}$. The surrounding $Z_d$ rotations set the required phase convention and may be implemented virtually on platforms with frame-update phase control.}
    \label{fig:diffusion}
\end{figure}

In Fig.~\ref{fig:diffusion}, we show a circuit identity for diffusion operators in which the $\mathtt{CPHASE}$ gate acts on an arbitrary basis state $\ket{m}$ rather than the standard state $\ket{0}$. The surrounding $Z_d$ rotations shift the phase convention using the identity $(Z_d^\dagger)^{d-m}=Z_d^m$.

\section{Proof of identities}
\label{app:identities}
In this section, we prove several qudit identities used in the main text. We begin with $H_d X_d^j = Z_d^j H_d$. First, note that
\begin{equation}
H_d^\dagger \;=\; \frac{1}{\sqrt{d}}
\sum_{j=0}^{d-1} \sum_{k=0}^{d-1}
\omega^{-jk} \, |j\rangle \langle k|,
\qquad
\omega = e^{2\pi i / d}.
\end{equation}
We start by applying $H_d^\dagger$ to an arbitrary basis state $\ket{k}$:
\begin{equation}
H_d^\dagger \ket{k}
=  \frac{1}{\sqrt d} \sum_{m=0}^{d-1} \omega^{-mk} \ket{m}.
\end{equation}

\begin{align}
X_d H_d^\dagger \ket{k}
&= X_d \left( \frac{1}{\sqrt d} \sum_{m=0}^{d-1} \omega^{-mk} \ket{m} \right) \\
&= \frac{1}{\sqrt d} \sum_{m=0}^{d-1} \omega^{-mk} \ket{m+1}.
\end{align}

Relabeling the summation index as $n=m+1$ modulo $d$, we obtain
\begin{align}
X_d H_d^\dagger \ket{k}
&= \frac{1}{\sqrt d} \sum_{n=0}^{d-1} \omega^{-(n-1)k} \ket{n} \\
&= \omega^{k} \frac{1}{\sqrt d} \sum_{n=0}^{d-1} \omega^{-nk} \ket{n}.
\end{align}

Applying the Hadamard (Fourier transform) operator,
\begin{align}
H_d X_d H_d^\dagger \ket{k}
&= \omega^{k} H_d \left( \frac{1}{\sqrt d} \sum_{n=0}^{d-1} \omega^{-nk} \ket{n} \right) \\
&= \omega^{k} \frac{1}{d} \sum_{j=0}^{d-1} \sum_{n=0}^{d-1}
\omega^{jn} \omega^{-nk} \ket{j}.
\end{align}
The sum over $n$ evaluates to
$
\sum_{n=0}^{d-1} \omega^{n(j-k)} = d\,\delta_{j,k},
$
and therefore
\begin{align}
H_d X_d H_d^\dagger \ket{k}
&= \omega^{k} \ket{k}.
\end{align}

Since $Z_d\ket{k}=\omega^{k}\ket{k}$, we conclude
\begin{equation}
H_d X_d H_d^\dagger = Z_d.
\end{equation}
One can use this expression recursively to show that
\begin{equation}
    H_d X_d^2 H_d^\dagger = (H_d X_d H_d^\dagger)(H_d X_d H_d^\dagger) = Z_d^2,
\end{equation}
and in general
\begin{equation}
    H_d X_d^j H_d^\dagger = Z_d^j \implies H_d X_d^j = Z_d^j H_d.
\end{equation}

Next, we show that $(Z_d^{\dagger})^{d-m} = (Z_d)^{m}$.

Raising $Z_d^\dagger$ to the $(d-m)$-th power gives
\begin{align}
\left(Z_d^\dagger\right)^{d-m}
&=
Z_d^{-(d-m)} \nonumber \\
&=
Z_d^{m-d} \nonumber \\
&=
Z_d^m Z_d^{-d} \nonumber \\
&=
Z_d^m,
\end{align}
since $Z_d^d=\mathbb{I}_{d}$.

\section{Gate complexity of multi-controlled \texorpdfstring{$\mathtt{CPHASE}(\theta)$}{CPHASE(theta)} decomposition}
\label{app:recursion}
The recursive construction in Fig.~\ref{fig:cphase}(a) implies that the two-qudit gate count $N_2$ satisfies
\begin{align}
    N_2(n) &= (d+1)N_2(n-1) + N_2(n-2) + d, \nonumber \\
    N_2(0) &= N_2(1) = 0.
\end{align}
The first term accounts for the $d+1$ applications of a $\mathtt{CPHASE}$ gate with one fewer control qudit, the second accounts for a $\mathtt{CPHASE}$ gate with two fewer control qudits, and the final term accounts for the $d$ two-qudit gates.

To determine the leading asymptotic behavior, we omit the inhomogeneous term $d$, which does not change the exponential growth rate. The resulting characteristic equation is
\begin{equation}
    x^{n} = (d+1)x^{n-1} + x^{n-2} \implies x^2 = (d+1)x + 1,
\end{equation}
whose dominant root is
\begin{equation}
    x_{+} = \frac{(d+1)+\sqrt{(d+1)^2+4}}{2}.
\end{equation}
Thus, $N_2(n)=\mathcal{O}(x_{+}^{n})$. Since $x_{+}=d+1+\mathcal{O}(d^{-1})$ for large $d$, this scaling approaches $\mathcal{O}((d+1)^n)$. Table~\ref{tab:gate_counts} lists the gate counts for $d=2$, $3$, and $5$. With ancilla qudits, however, the gate count may be reduced to $\mathcal{O}(nd)$.
\begin{table}[H]
\centering
\caption{Numbers of two-qudit ($N_2$) and single-qudit ($N_1$) gates required to decompose a multi-controlled qudit $\mathtt{CPHASE}$ gate as functions of the number of qudits $n$, for qubit ($d=2$), qutrit ($d=3$), and ququint ($d=5$) systems.}
\label{tab:gate_counts}
\setlength{\tabcolsep}{8pt}
\begin{tabular}{c cc cc cc}
\toprule
& \multicolumn{2}{c}{$d=2$}
& \multicolumn{2}{c}{$d=3$}
& \multicolumn{2}{c}{$d=5$} \\
\cmidrule(lr){2-3}\cmidrule(lr){4-5}\cmidrule(l){6-7}
$n$ & $N_2$ & $N_1$ & $N_2$ & $N_1$ & $N_2$ & $N_1$ \\
\midrule
1 &    0 &    1 &    0 &    1 &    0 &    1 \\
2 &    2 &    3 &    3 &    4 &    5 &    6 \\
3 &    8 &   10 &   15 &   17 &   35 &   37 \\
4 &   28 &   33 &   66 &   72 &  220 &  228 \\
5 &   94 &  109 &  282 &  305 & 1360 & 1405 \\
6 &  312 &  360 & 1197 & 1292 & 8385 & 8658 \\
\bottomrule
\end{tabular}
\end{table}

\section{Equations for phase parameters}
\label{app:equations}

The following two equations determine the phases for the D2p protocol when the chosen iteration count $k_{\rm odd}\ge k_{\rm certain}$ is odd.
\begin{widetext}
\begin{subequations}
	\begin{align}
	\begin{split}
		2\lambda &+ (1-2\lambda)\cos(\beta_1) \\
		&- (1-2\lambda)\sin\left( \dfrac{\beta_1}{2} \right)
		\Bigl[ \sin(\beta_1) \cos\left( \dfrac{\beta_2}{2} \right) \\
		&\qquad + \Big(1+4\lambda-8\lambda^2 + (1-8\lambda+8\lambda^2)\cos(\beta_1)\Big) \sin\left( \dfrac{\beta_2}{2} \right) \Bigr]
		\dfrac{\tan(\frac{k_{\rm odd}-1}{2} \phi)}{\sin(\phi)} = 0, 
		\end{split}
		\\
		\begin{split}
		(1 &- 2\lambda)\sin(\beta_1) \\
		&+\Bigl[ (1-2\lambda)\Bigl(
		8\lambda(1-\lambda)\sin\left( \dfrac{\beta_1}{2} \right) \sin(\beta_1)\sin\left( \dfrac{\beta_2}{2} \right) \\
		&\qquad +
		\cos(\beta_1)\sin\left( \dfrac{\beta_1+\beta_2}{2} \right) \Bigr) -
		2\lambda\sin\left( \dfrac{\beta_1-\beta_2}{2} \right)\Bigr]
		\dfrac{\tan(\frac{k_{\rm odd}-1}{2} \phi)}{\sin(\phi)} = 0.
		\end{split}
	\end{align}
\end{subequations}
\end{widetext}

The following equations apply to the D3p protocol when only the final two steps have non-$\pi$ diffusion phases, i.e., $k'=k_{\rm certain}-2$ in Eq.~\eqref{eq:d3p}. In this case,
$\ket{T} = G(\pi,\beta_2)\cdot G(\pi,\beta_1)\cdot [G(\pi,\pi)]^{k_{\rm certain}-2} \ket{\psi_0}$.

\begin{widetext}
\begin{subequations}
\begin{align}
\lambda(1 &-2\lambda) - 2\lambda(1-\lambda)
\Bigl(\cos\beta_1 + \cos\beta_2 \Bigr)
- \bigl[1-\lambda(3-2\lambda)\bigr]\cos(\beta_1+\beta_2)
\nonumber\\
&\quad
+ \sqrt{\lambda(1-\lambda)}
\Bigl(2\lambda\cos\beta_1
- 2(1-\lambda)\cos\beta_2
+ (1-2\lambda)\bigl[1+\cos(\beta_1+\beta_2)\bigr]\Bigr)
\tan\!\left[\left(2k_{\rm certain} - 3 \right) \sin^{-1}(\sqrt{\lambda})\right] = 0,
\\ 
(1 &-\lambda)
\Bigl(2\lambda(\sin\beta_1+\sin\beta_2)
+ (1-2\lambda)\sin(\beta_1+\beta_2)\Bigr)
\nonumber\\
&\quad
- \sqrt{\lambda(1-\lambda)}
\Bigl(2\lambda\sin\beta_1
- 2(1-\lambda)\sin\beta_2
+ (1-2\lambda)\sin(\beta_1+\beta_2)\Bigr)
\tan\!\left[\left(2k_{\rm certain} - 3 \right) \sin^{-1}(\sqrt{\lambda})\right] = 0.
\end{align}
\end{subequations}
\end{widetext}
These equations for D2p and D3p can be solved numerically with the D1p phase as an initial guess,
\begin{equation}
    \beta_{\rm guess} = 2 \sin^{-1} \left( \frac{1}{\sqrt{\lambda}} \sin{\left(\frac{\pi}{4k_{\rm certain}+2} \right)} \right).
\end{equation}

Because the phases are $2\pi$-periodic, solutions that differ by integer multiples of $2\pi$ are equivalent. For the D3p example in the main text, $\lambda=0.005$ and $k_{\rm certain}=11$, and the final two phase values are $\beta_1=2.663$ and $\beta_2=-1.533$ radians.

\section{Properties of Chebyshev functions}
\label{app:cheby}

For an integer $n\ge0$, the $n$th Chebyshev polynomial of the first kind is defined as
\begin{equation}
T_n(x)=
\begin{cases}
\cos\!\left(n\,\arccos x\right), & |x|\le 1, \\[6pt]
\cosh\!\left(n\,\operatorname{arccosh}x\right), & x\ge 1, \\[6pt]
(-1)^n\cosh\!\left(n\,\operatorname{arccosh}(-x)\right), & x\le -1.
\end{cases}
\end{equation}
An equivalent algebraic expression is
\begin{equation}
    T_n(x) = \frac{1}{2}\left[ (x+\sqrt{x^2-1})^n + (x-\sqrt{x^2-1})^n \right].
\end{equation}
This identity is valid for all complex $x$ with appropriate branch choices. For the noninteger order used in the YLC protocol, only the $x\ge1$ branch is required, and $T_\nu(x)=\cosh[\nu\operatorname{arccosh}(x)]$. The Chebyshev functions satisfy the recurrence relation
\begin{equation}
    T_{\nu+2}(x) = 2x\,T_{\nu+1}(x) - T_{\nu}(x).
\end{equation}






\bibliography{main}

\end{document}